\documentclass[a4paper]{jpconf}
\usepackage{graphicx}
\begin{document}
\title{Spin dependence of the antinucleon-nucleon interaction}

\author{Johann Haidenbauer}

\address{
Institute for Advanced Simulation and J\"ulich Center for Hadron Physics,
Forschungszentrum J\"ulich, D-52425 J\"ulich, Germany}

\ead{j.haidenbauer@fz-juelich.de}

\begin{abstract}
The status of our present knowledge on the 
antinucleon-nucleon interaction at low and medium
energies is discussed. Special emphasis is put on
aspects related to its spin dependence which are relevant 
for experiments planned by the PAX collaboration. 
Predictions for the spin-dependent $\bar pp$ cross 
sections $\sigma_1$ and $\sigma_2$ are presented, 
utilizing $\bar NN$ potential models developed by the
J\"ulich group, and compared to results based on the
amplitudes of the Nijmegen partial-wave analysis. 
\end{abstract}

\section{Introduction}

While the interest in antinucleon physics waned somewhat 
after 1996 when the Low Energy Anti-Proton Ring (LEAR) at CERN 
was shut down, this trend has clearly reversed over the last
couple of years. 
Hereby measurements of the antiproton-proton ($\bar pp$) invariant mass 
in the decays of $J/\psi$, $B$ mesons, etc., definitely played an
important role where in some of those studies a near-threshold enhancement
in the mass spectrum was found \cite{Bai,Aubert,Aubert1}. 
In case of the radiative decay of $J/\psi$ this enhancement turned out to 
be so spectacular that it even nourished 
speculations that one might have found evidence for the existence 
of $\bar pp$ bound states \cite{Bai,Dedonder}. 
The most important factor is certainly the proposed Facility for Antiproton and 
Ion Research (FAIR) in Darmstadt whose construction is finally on its
way. Among the various experiments planned at this site is the PANDA Project 
\cite{PANDA} which aims to study the interactions between antiprotons
and fixed target protons and nuclei in the momentum range of
1.5-15 GeV/c using the high energy storage ring HESR. 

Another project suggested for the FAIR facility comes from the PAX 
collaboration. This collaboration was formed \cite{PAX} with the aim
to measure the proton transversity in the interaction of polarized
antiprotons with protons.
In order to produce an intense beam of polarized antiprotons, the collaboration
intends to use antiproton elastic scattering off a polarized hydrogen
target ($^1$H) in a storage ring \cite{Rathmann}.
The basic idea is connected to the result of the FILTEX experiment \cite{FILTEX},
where a sizeable effect of polarization buildup was achieved in a
storage ring by scattering of unpolarized protons 
off a polarized hydrogen atoms at low beam energies of 23 MeV. 
Recent theoretical analyses \cite{MS,NNNP,NNNP1,NNNP2} have shown that
the polarization buildup observed in Ref.~\cite{FILTEX} can be understood
quantitatively. According to those authors it is solely due to the spin 
dependence of the hadronic (proton-proton) interaction which leads to the
so-called spin-filtering mechanism, i.e. to a different rate of removal
of beam protons from the ring for different polarization states of the
target proton.  

In contrast to the $NN$ case, the spin dependence of the $\bar NN$
interaction is poorly known. Therefore, it is an open question whether 
any sizeable polarization buildup can also be achieved in case of an
antiproton beam based on the spin-filtering mechanism. Indeed, 
recently several theoretical studies were performed with the aim
to estimate the expected polarization effects for antiprotons,
employing different $\bar{p}p$ interactions 
\cite{DmitrievMS,Uzikov,Salnikov,Salnikov1}. 
In this contribution I provide an overview and a comparison of the
results of those investigations. Furthermore, I take the opportunity
to briefly recall the status of our knowledge of the spin dependence 
of the $\bar NN$. Thereby I focus on the only double-polarization
observables measured so far, namely the depolarization parameter 
$D_{NN}$ and the spin parameter $K_{NN}$ \cite{DNEL,DNCE,DNCE1,KNCE}

Besides of using polarized protons as target one could also use 
light nuclei as possible source for the antiproton polarization
buildup. A corresponding investigation for antiproton scattering
on deuterons was presented in Ref.~\cite{Uzikov,Uzikov1}, cf. also a
related contribution to these proceedings \cite{Uzikov2}.

\begin{figure}
\includegraphics{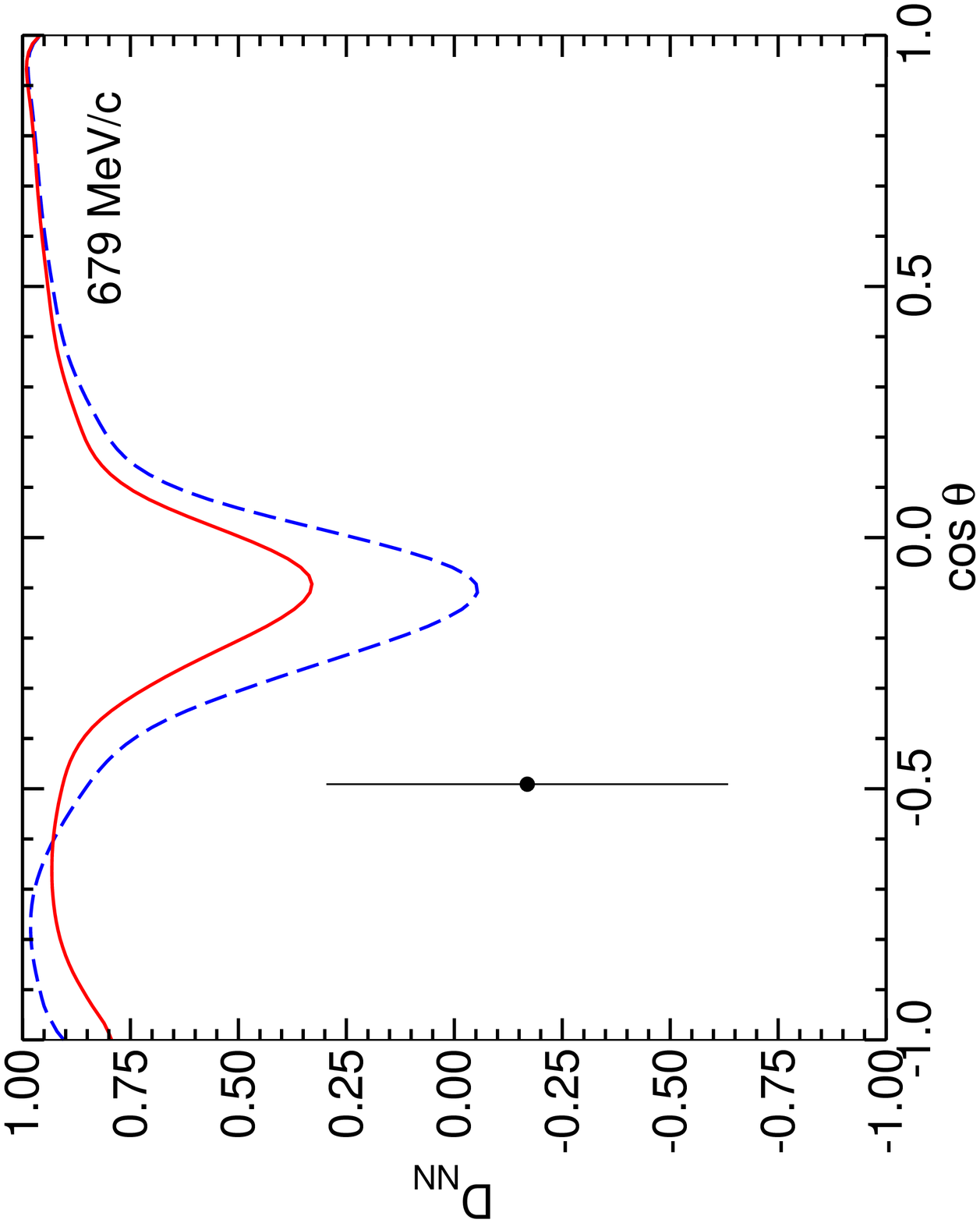}
\includegraphics{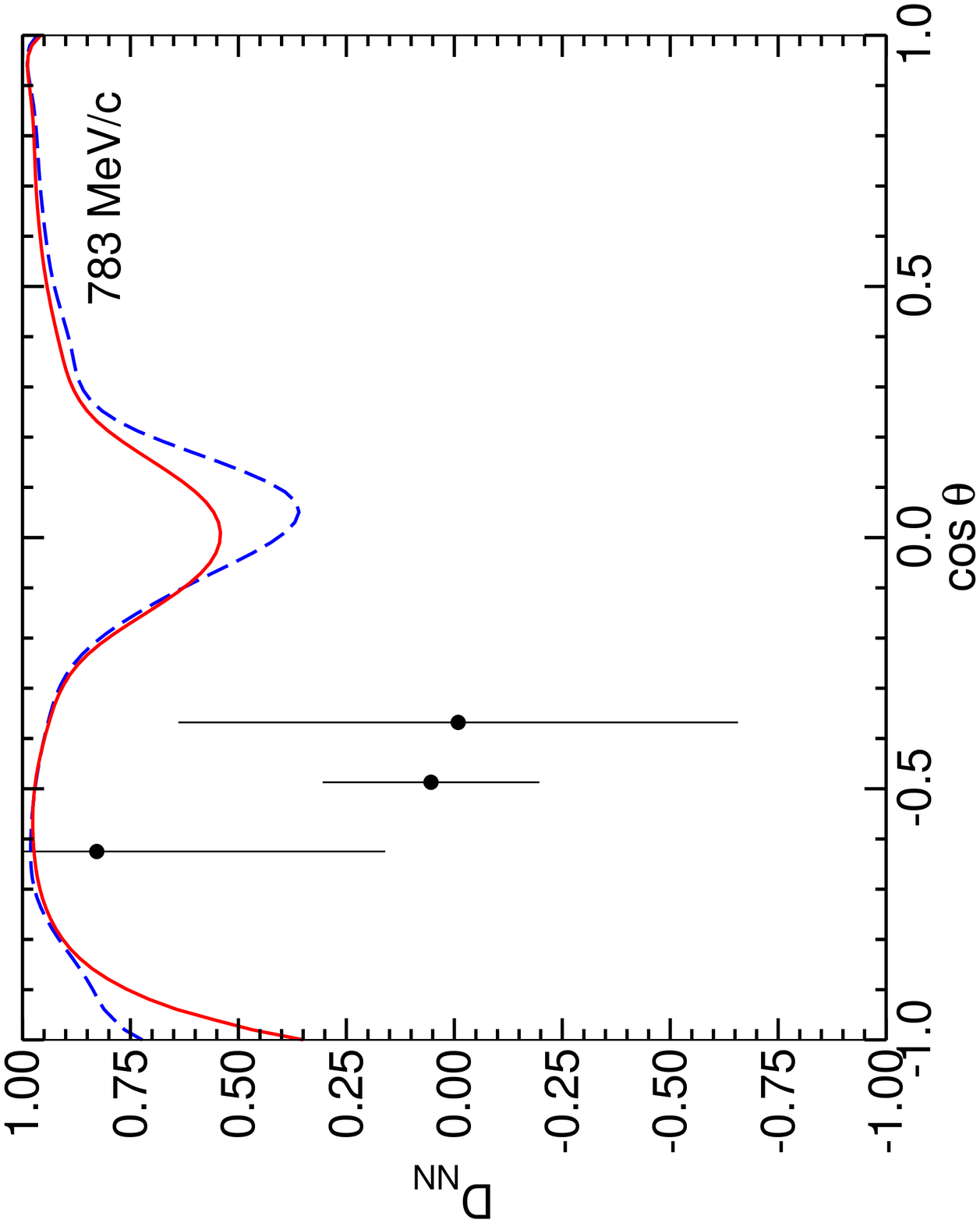}
\includegraphics{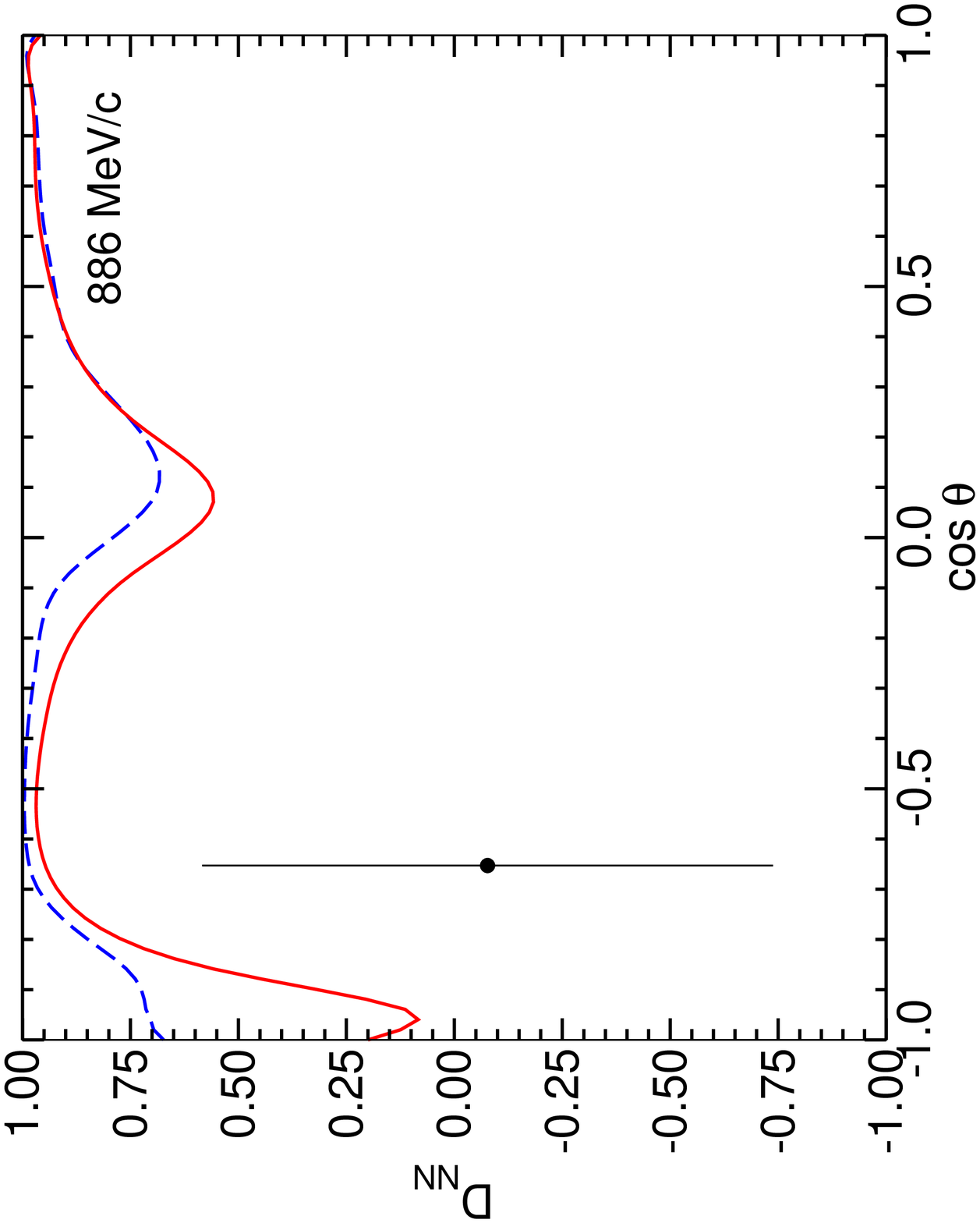}
\includegraphics{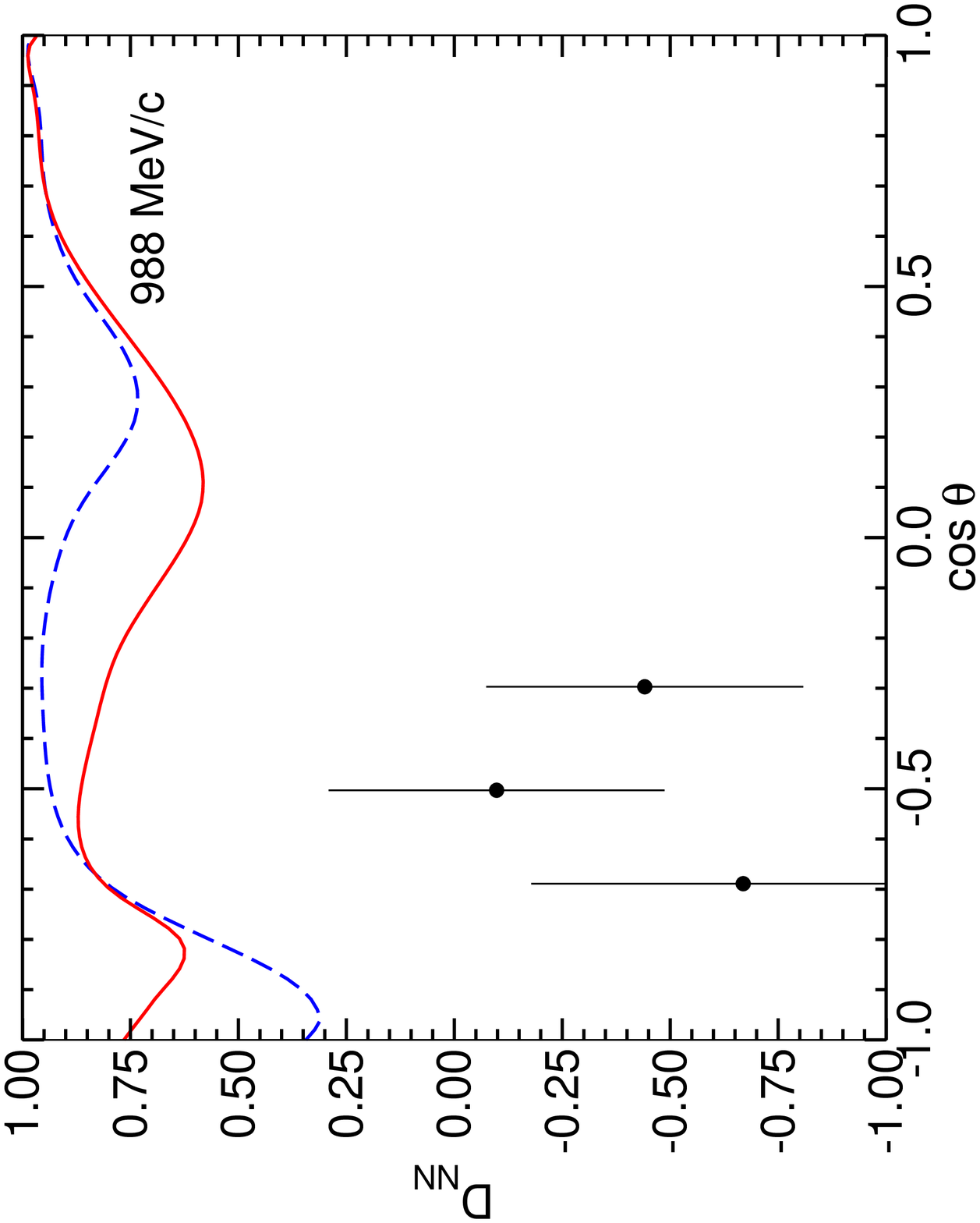}
\vskip 10.5cm
\caption{Depolarization parameter $D_{NN}$ for $\bar p p \to \bar p p$
at selected beam momenta $p_{lab}$. The data are taken from 
Ref.~\cite{DNEL}. 
Results for model D (solid line) and model A (dashed line)
are shown.
}
\label{dnel}
\end{figure}

\section{The J\"ulich $\bar NN$ models}

Since most of the results reported below are for $\bar NN$
interactions developed by the J\"ulich group let me briefly summarize 
the salient features of those potential models. In fact,  
the J\"ulich group has developed several $N\bar N$ models over the
years \cite{Hippchen0,Hippchen,Hippchen1,Hai1,Mull}. 
In the following I am going to present results for the models 
A(BOX), introduced in Ref.~\cite{Hippchen}, and D, described in Ref.~\cite{Mull}.
Starting point for both models is the full Bonn $NN$ potential~\cite{MHE};
it includes not only traditional one-boson-exchange diagrams but also
explicit $2\pi$- and $\pi\rho$-exchange processes as well as virtual
$\Delta$-excitations. The G-parity transform of this meson-exchange
$NN$ model provides the elastic part of the considered $N\bar N$ interaction
models.
In case of model A(BOX) \cite{Hippchen} (in the following
referred to as model A)
a phenomenological \hbox{spin-}, isospin- and energy-independent
complex potential of Gaussian form is added to account for the
$N\bar N$ annihilation. It contains only three free parameters 
(the strengths of the real and imaginary parts of the annihilation
potential and its range), fixed in a fit to the available total and integrated
$\bar NN$ cross sections.
In case of model D \cite{Mull}, the most complete $N\bar N$ model of the
J\"ulich group, the $N\bar N$ annihilation into 2-meson decay
channels is described microscopically, including all possible
combinations of $\pi$, $\rho$, $\omega$, $a_0$, $f_0$, $a_1$, $f_1$,
$a_2$, $f_2$, $K$, $K^+$ -- see Ref. \cite{Mull} for details --
and only the decay into multi-meson channels is simulated by
a phenomenological optical potential.

\begin{figure}
\includegraphics{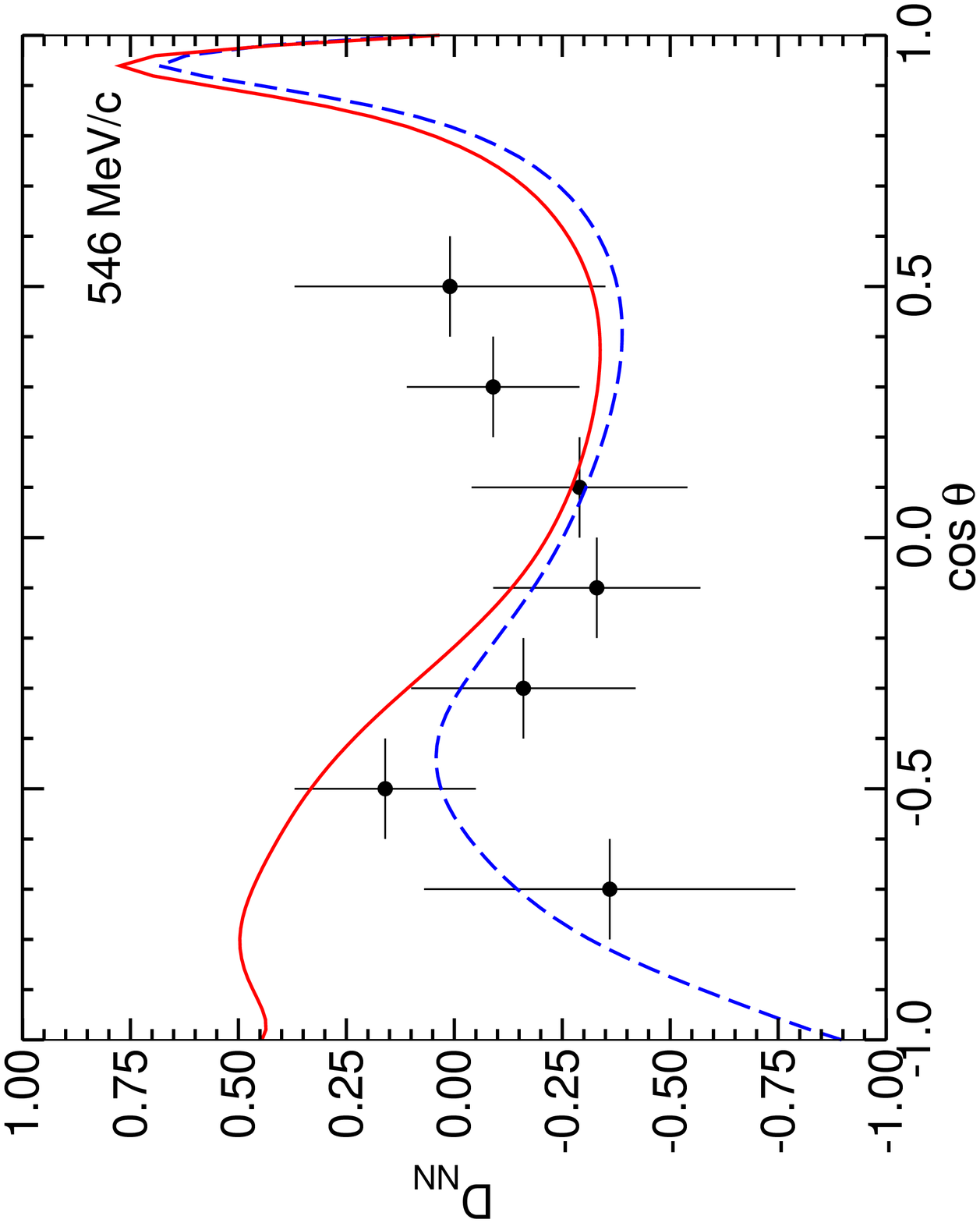}
\includegraphics{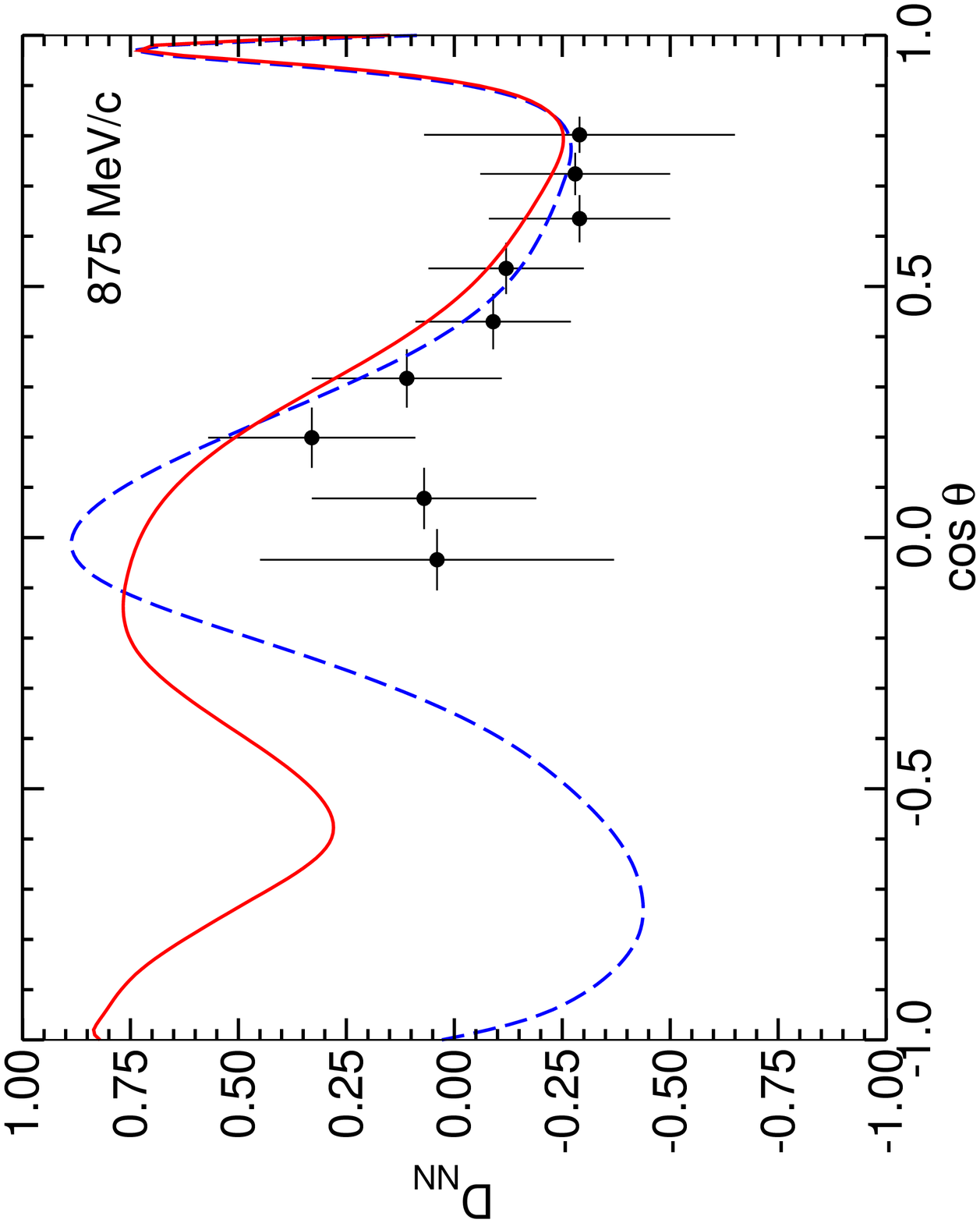}
\includegraphics{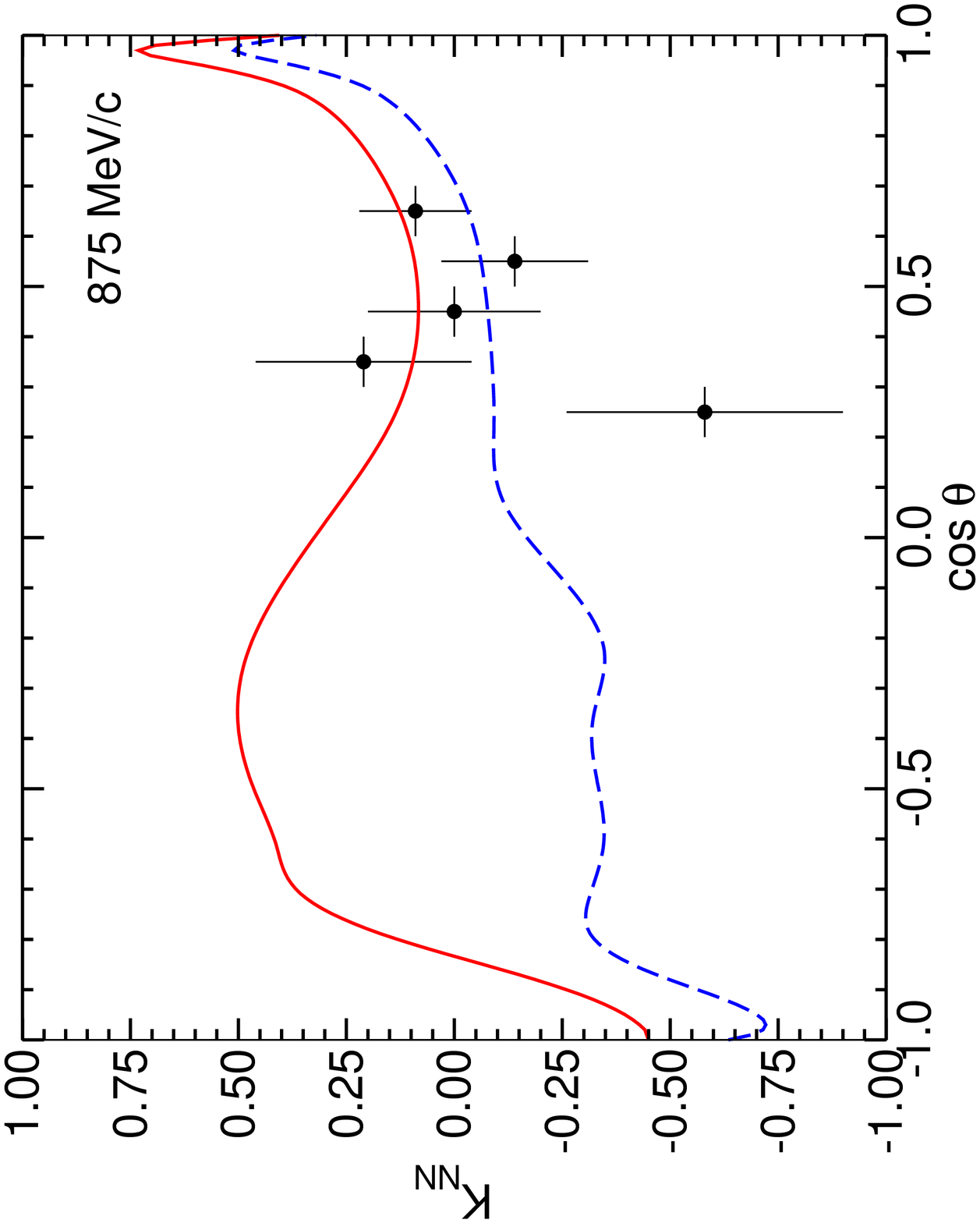}
\vskip 10.5cm
\caption{Double-polarization observables $D_{NN}$ and $K_{NN}$
for $\bar p p \to \bar n n$ at selected beam momenta $p_{lab}$. 
The data are taken from Refs.~\cite{DNCE,DNCE1,KNCE}.
Results for model D (solid line) and model A (dashed line)
are shown.
}
\label{dnce}
\end{figure}

\section{Results}

Results for the total and integrated elastic ($\bar p p$) and
charge-exchange ($\bar p p \to \bar nn$) cross sections and also
for angular dependent observables for both models can be found in
Refs.~\cite{Hippchen,Mull}. As one can see there, with model A as 
well as with D a very good overall reproduction of the low- and 
intermediate energy $\bar N N$ data is achieved.
Moreover, exclusive data on several $p\bar p$ 2-meson and even
3-meson decay channels are described with fair quality
\cite{Hippchen1,Mull,Betz}.
Recently it has been shown that the
$\bar NN$ models of the J\"ulich group can also explain successfully
the near-threshold enhancement seen in the $\bar pp$ mass
spectrum of the reactions $J/\psi \to \gamma \bar pp$ \cite{Sibi1},
$J/\psi \to \omega\bar pp$ \cite{Sibi3} and $B^+ \to K^+ \bar pp$
\cite{Sibi2} and in the $e^+e^-\to \bar pp$
cross section \cite{Sibi4}.

In Ref.~\cite{Uzikov} we presented a comparison of the 
J\"ulich $\bar NN$ interaction 
with existing data for the $\bar n p$ channel \cite{Arm87,Iaz00},
which is a purely isospin $I=1$ system. It showed that 
the J\"ulich models are in nice agreement with the experimental
information on the $\bar n p$ interaction too, despite the fact
that those data on total and annihilation cross sections have 
not been included in the fitting procedure.

\begin{table}[h]
\renewcommand{\arraystretch}{1.10}
\caption{Partial cross sections predicted by the J\"ulich 
$\bar NN$ models A and D \cite{Hippchen,Mull} in comparison 
to results from the Nijmegen $\bar NN$ partial wave analysis 
\cite{Timmermans}.} 
\label{table1}
\centering
\begin{tabular}{|c|c|rrrr|rrrr|}
\hline
&& \multicolumn{4}{|c|}{$\bar p p \to \bar p p$} & \multicolumn{4}{|c|}{$\bar p p \to \bar n n$} \\
& $p_{lab}$ (MeV/c) & 200 & 400 & 600& 800 & 200 & 400 & 600& 800 \\
\hline
$^1S_0$ & {A}   & 15.0& 7.7& 4.4& 2.7 & 0.8 & 0.1 & &  \\
        & {D}   & 12.9& 7.8& 4.9& 3.3 & 0.4 &     & &  \\
        & {Nijmegen} & 14.6& 6.8& 3.7& 1.9 & 0.5 & & &  \\
\hline
$^3S_1$ & {A}   & 72.8& 30.4& 14.5& 7.9 & 0.6 & 0.2 & 0.1&     \\
        & {D}   & 68.2& 22.8&  9.8& 6.3 & 4.9 & 1.4 & 0.4& 0.1 \\
        & {Nijmegen} & 71.1& 29.6& 14.4& 7.9 & 2.0 & 0.5 & 0.3 & 0.3 \\
\hline
$^3P_0$ & {A}   &  3.3& 3.1& 2.5& 1.9 & 2.9 & 1.1 & 0.4 & 0.1 \\
        & {D}   &  2.0& 2.3& 2.4& 2.3 & 2.5 & 0.6 & 0.1 &  \\
        & {Nijmegen} &  4.7& 4.6& 3.4& 2.6 & 2.1 & 1.4 & 0.7 & 0.3 \\
\hline
$^1P_1$ & {A}   &  2.3& 5.2& 5.5& 4.9 & 1.1 & 0.4 & 0.2 & 0.1 \\
        & {D}   &  4.1& 7.5& 7.2& 6.0 & 0.7 & 0.8 & 0.5 & 0.3 \\
        & {Nijmegen} &  1.7& 2.6& 2.5& 2.4 & 1.3 & 0.4 &     &     \\
\hline
$^3P_1$ & {A}   &  3.8& 9.5& 8.1& 6.2 & 5.6 & 2.7 & 0.5 & 0.1 \\
        & {D}   &  4.6&10.4& 8.3& 5.8 & 4.9 & 2.4 & 0.7 & 0.2 \\
        & {Nijmegen} &  1.8& 7.6& 7.6& 6.2 & 6.7 & 6.2 & 2.8 & 1.4 \\
\hline
$^3P_2$ & {A}   &  4.9&14.8&13.4&10.1 & 0.6 & 0.7 & 0.2 & 0.1 \\
        & {D}   &  4.9&14.7&14.3&11.5 & 0.6 & 0.6 & 0.2 & 0.1 \\
        & {Nijmegen} &  6.3&16.1&15.5&12.9 & 0.8 & 1.2 & 0.5 & 0.3 \\
\hline
\end{tabular}
\end{table}

As already mentioned in the Introduction, the spin dependence of the
$\bar NN$ interaction is not well known. There is a fair amount of
data on analyzing powers, for $\bar pp$ elastic as well as for
$\bar pp \to \bar nn$ charge-exchange scattering, cf.
Ref.~\cite{Klempt} for a recent review.
The predictions
of the J\"ulich models A and D are in reasonable agreement with
the experimental polarizations up to beam momenta of
$p_{lab}\approx 550$ MeV/c as can be seen in Ref.~\cite{Mull}.
In fact, model A gives a somewhat better account of the data and
reproduces the measured $\bar pp$ polarizations even
quantitatively up to $p_{lab}\approx 800$ MeV/c
($T_{lab} \approx 300$ MeV).
 
As far as other spin-dependent observables are concerned, 
specifically with regard to double-polarization observables, 
there is only scant information on the depolarization $D_{NN}$
\cite{DNEL,DNCE,DNCE1} and also on the spin parameter $K_{NN}$ \cite{KNCE}. 
Moreover, those data are of rather limited
accuracy so that they do not really provide serious constraints 
on the $\bar NN$ interaction. In order to illustrate the present
status I show here those data together with predictions of the
J\"ulich models. They can be found in Fig.~\ref{dnel} (for 
$\bar pp \to \bar pp$) and in Fig.~\ref{dnce} (for $\bar pp \to \bar nn$).
Interestingly the model results for the charge-exchange reaction
are more or less in line with the data, while the depolarization
predicted for elastic scattering is certainly too large. For
fairness one should say that some of those data are already outside
of the momentum range ($p_{lab} \le 800$ MeV/c) for which the 
J\"ulich $\bar NN$ models were originally designed \cite{Hippchen,Mull}. 

In this context let us mention that a partial-wave analysis (PWA)
of $\bar pp$ scattering has been performed by the Nijmegen
Group \cite{Timmermans} which, in principle, would allow to pin
down the spin-dependence of the $\bar NN$ interaction. However,
the uniqueness of the achieved solution was disputed in Ref.~\cite{Richard}.
Moreover, the actual partial-wave amplitudes of the Nijmegen analysis
are not readily available and, therefore, one cannot confront the
results of the J\"ulich models directly with those of the 
Nijmegen PWA. But at least the authors of \cite{Timmermans} published 
partial elastic and charge-exchange cross sections. 
In Table \ref{table1} the results of the J\"ulich $\bar NN$ models 
A and D are compared with those of the Nijmegen PWA for the
$S$- and $P$ waves. 

As one can see from the Table, qualitatively there is a good overall 
agreement
between the two models and the Nijmegen analysis. This may be not too
surprising in view of the fact that all of them reproduce the bulk
properties of $\bar pp$ scattering rather well. But one can see also
drastic quantitative differences, specifically in the $P$ waves, 
where in some cases the partial cross sections of the Nijmegen
analysis differ by factors of 2 or even more from those of the 
J\"ulich $\bar NN$ interactions. There are noticeable differences between
the predictions of the models as well. Clearly those differences will
be reflected in the results for the spin-dependent $\bar pp$ cross 
sections which are considered next.

The total polarized $\bar pp$ cross section can be written as
\begin{equation}
\nonumber
\sigma_{tot} = \sigma_{0}
+ \sigma_{1} ({\bf P}_B \cdot {\bf P}_T)
+ \sigma_{2} ({\bf P}_B \cdot {\bf \hat{k}}) ({\bf P}_T \cdot {\bf \hat{k}})
\end{equation}
where ${\bf P}_B$ and ${\bf P}_T$ are the polarization vectors of the
beam and target, respectively, and ${\bf \hat{k}}$ is a unit vector in the
direction of the beam \cite{Bystricky}. In terms of the standard
helicity amplitudes $M_i(\theta)$ ($i=1,...,5$) the cross sections 
are given by \cite{Bystricky} 
\begin{eqnarray}
\nonumber
\sigma_{0} &=& \phantom{-} \frac{2\pi}{k} {\rm Im} [M_1(0) + M_3(0)] \\
\nonumber
\sigma_{1} &=& \phantom{-} \frac{2\pi}{k} {\rm Im} [M_2(0)]  \\
\sigma_{2} &=& - \frac{2\pi}{k} {\rm Im} [M_1(0) + M_2(0) - M_3(0)], 
\label{CNpp2} 
\end{eqnarray}
where $k$ is the modulus of the center-of-mass momentum of the antiproton. 
Note that the above equations are only valid for the purely hadronic contribution
(called $\sigma_i^h$ in the following) to the cross sections.  
The Coulomb-nuclear interference contribution to the cross sections, 
$\sigma_i^{int}$, has to be calculated by integration
of the polarized differential $\bar pp$ cross section (with the Coulomb amplitudes
included in the reaction amplitude) over the scattering angle within the interval
$[\theta_{acc},\pi]$, where $\theta_{acc}$ is the beam acceptance angle 
\cite{MS,Uzikov}. Then the total spin-dependent cross sections $\sigma_i$ 
($i=1,2$) are given by the sum $\sigma_i^h+\sigma_i^{int}$

\begin{figure}
\includegraphics{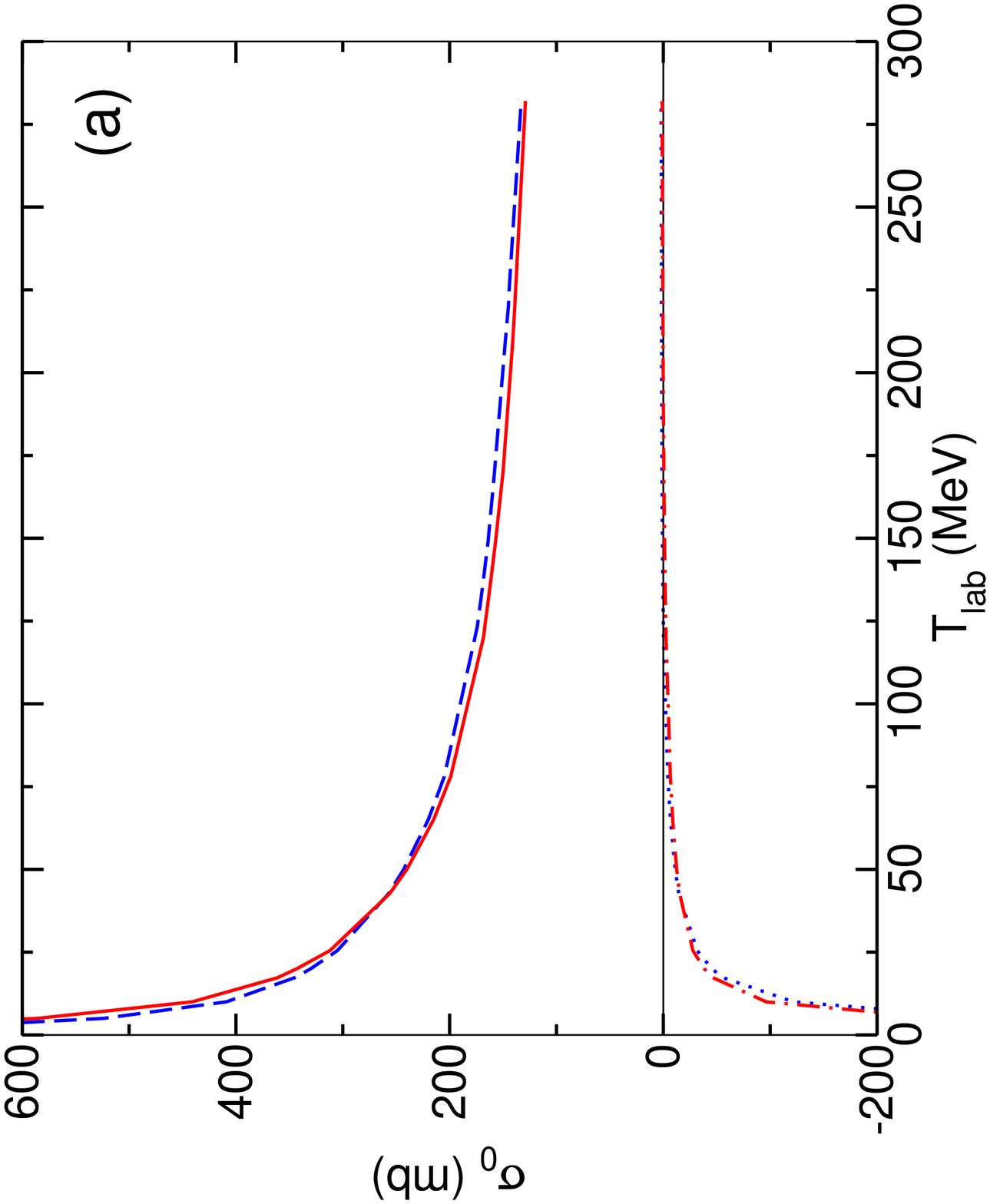}
\includegraphics{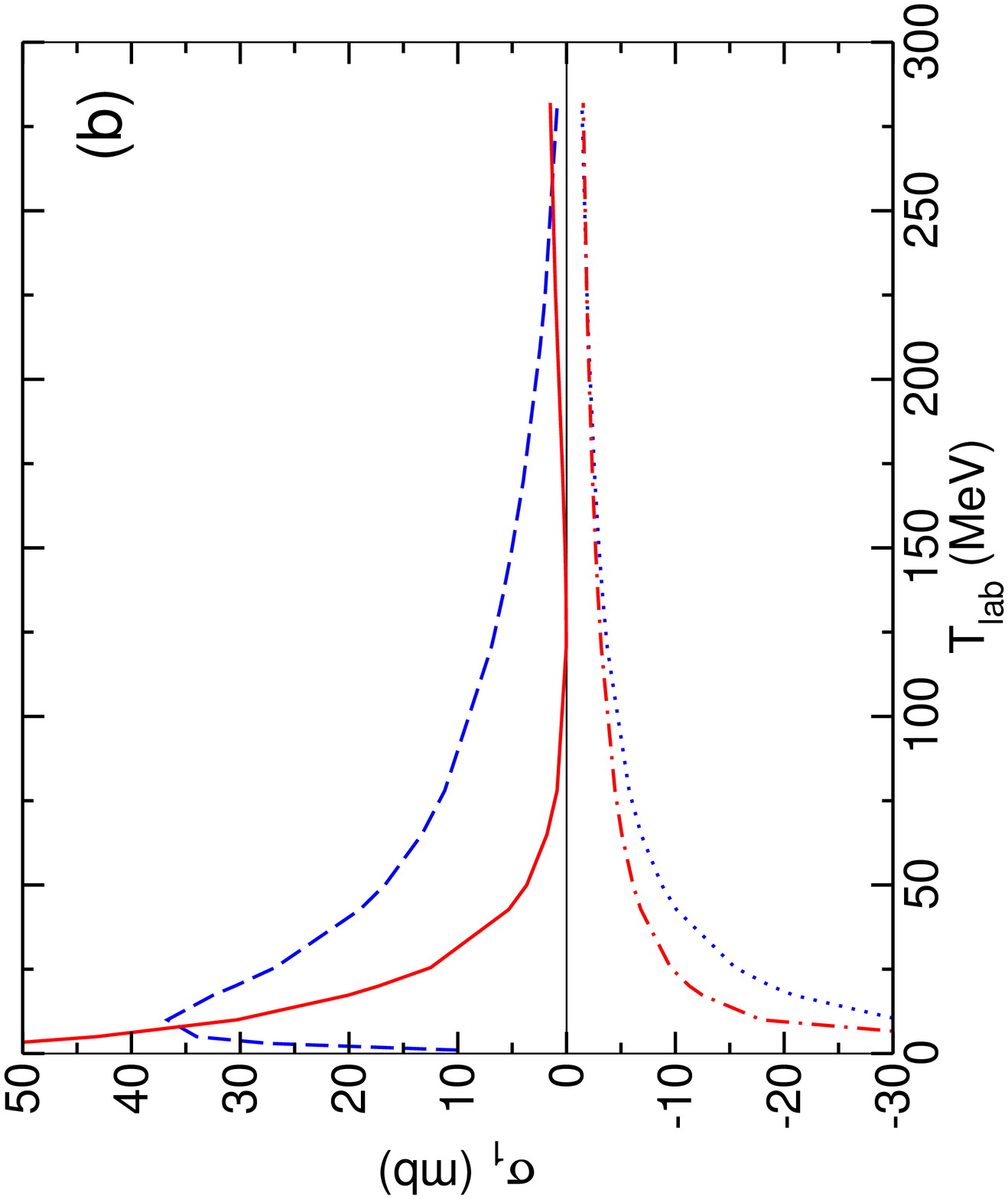}
\includegraphics{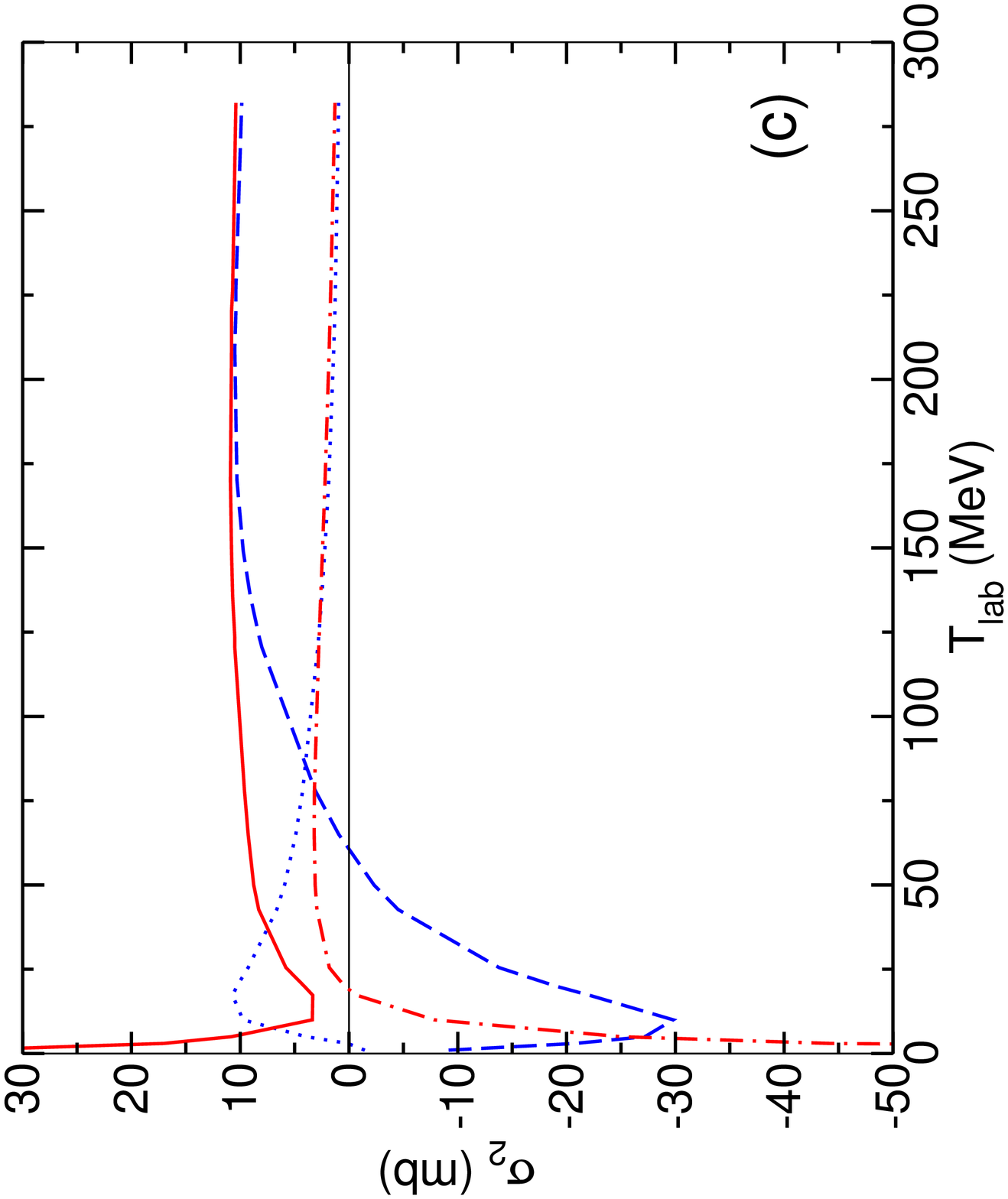}
\vskip 10.5cm
\caption{Total $\bar p p$ cross sections $\sigma_{0}$ (a),
$\sigma_{1}$ (b), and $\sigma_{2}$ (c) versus antiproton laboratory 
energy $T_{lab}$. 
Results based on the purely hadronic amplitude ($\sigma^h_i$) of the
models D (solid line) and A (dashed line) are shown, together with
those for the Coulomb-nuclear interference term ($\sigma^{int}_i$), 
corresponding to the dash-dotted (D) and dotted (A) lines, respectively.
}
\label{totppDA}
\end{figure}
\begin{figure}
\includegraphics{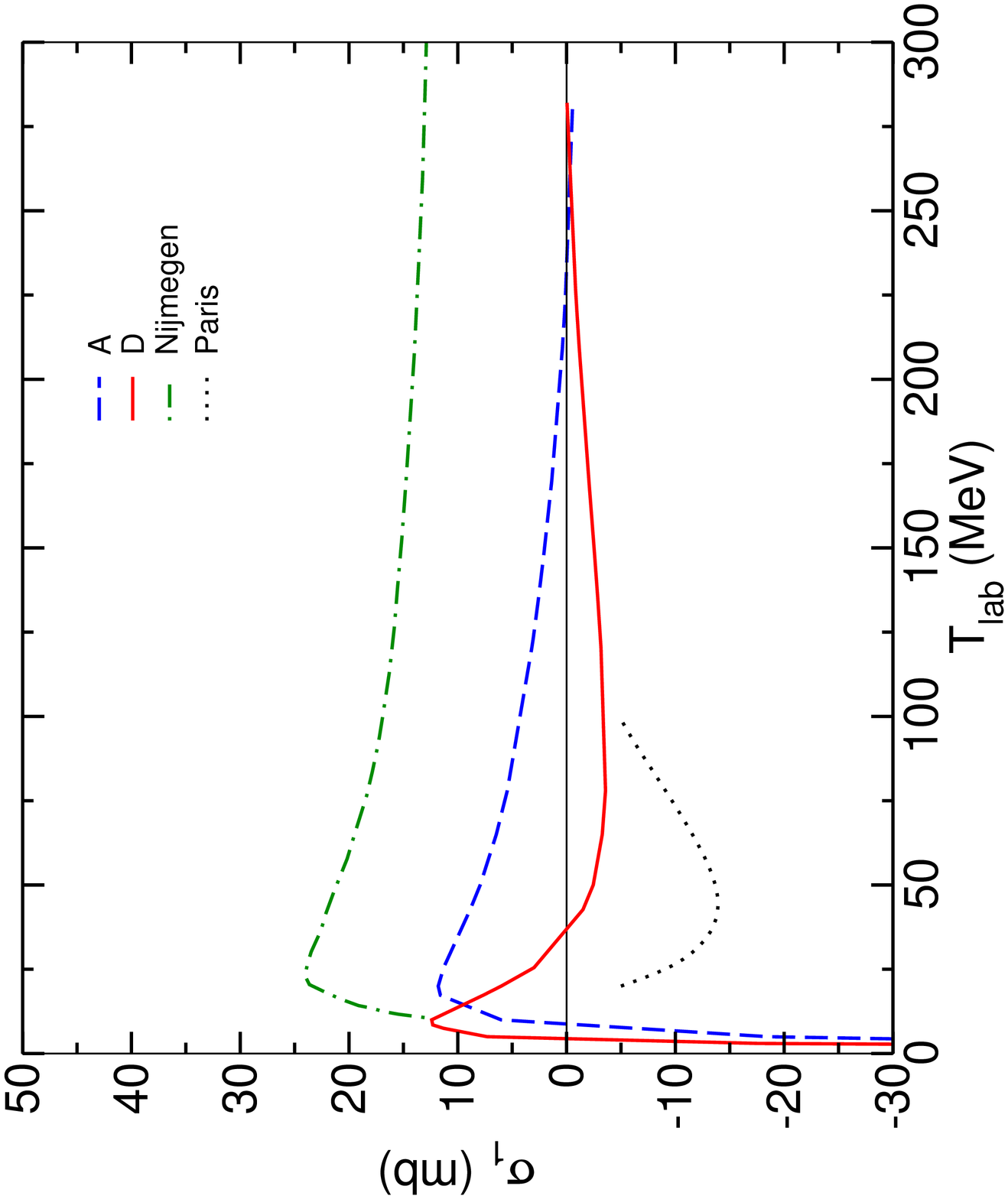}
\includegraphics{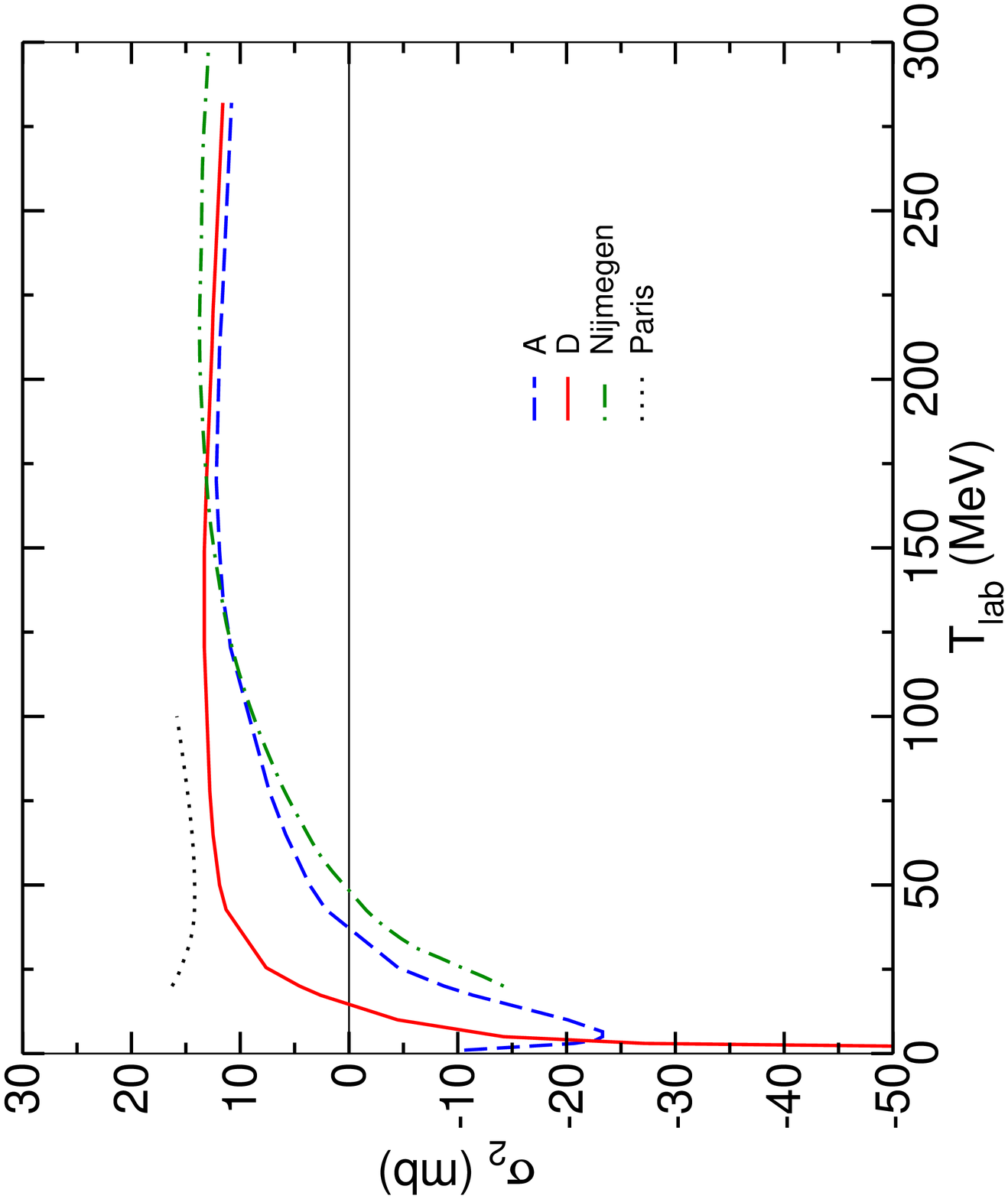}
\vskip 5.5cm
\caption{Spin-dependent $\bar p p$ cross sections $\sigma_{1}$ and 
$\sigma_{2}$ versus antiproton laboratory energy $T_{lab}$. 
Predictions for the total cross section ($\sigma_i = \sigma^{h}_i+\sigma^{int}_i$) 
of the J\"ulich models D (solid line) and A (dashed line) are shown. 
Corresponding results for the Nijmegen PWA (dash-dotted line) are taken from 
Ref.~\cite{Salnikov}, while those for the Paris potential (dotted line) are
from Ref.~\cite{DmitrievMS}. 
}
\label{totppC}
\end{figure}

Predictions of the J\"ulich $\bar NN$ interactions for the spin-dependent
$\bar pp$ cross sections are presented in Figs.~\ref{totppDA} and \ref{totppC}. 
Figure~\ref{totppDA} contains results based on the purely hadronic
amplitude ($\sigma_i^{h}$) and the Coulomb-nuclear interference term
($\sigma_i^{int}$) separately so that one can see the magnitude of the
latter. In the concrete calculations the acceptance angle was chosen to be
$\theta_{acc} = 8.8$ mrad \cite{FILTEX}.

At low energies, i.e. around $T_{lab} = 5\sim 10$ MeV, the
interference terms are comparable to the corresponding purely hadronic
cross sections and their magnitude increases further
with decreasing energy due to the $1/k$ factor, cf.
Eqs.~(27) in Ref.~\cite{Uzikov}.
With increasing energy the relevance of the Coulomb-nuclear interference
terms diminishes more and more in case of the cross sections $\sigma_0$
and $\sigma_2$. But for $\sigma_1$ the term is still significant, as one 
can see from Fig.~\ref{totppDA}b.
Note, that the three cross sections $\sigma_i^{int}$ ($i=0,1,2$)
themselves are all roughly of comparable magnitude for energies from
around 50 MeV onwards.

While the predictions of the two models for $\sigma_0$ are rather similar
(cf. Figs.~\ref{totppDA}a), even for the Coulomb-nuclear
interference cross section, this is not the case for the spin-dependent
cross sections $\sigma_1$ and $\sigma_2$. For energies below $T_{lab}
\approx 150$ MeV there are drastic differences between the results based
on the two models. Indeed, for $\sigma_2$ at low energies even the
sign differs in case of the $\bar pp$ channel. Obviously, here the variations
in the hadronic amplitude are also reflected in large differences in
the Coulomb-nuclear interference term.

In Fig.~\ref{totppC} the total spin-dependent cross sections $\sigma_1$ and 
$\sigma_2$ are displayed.
As far as $\sigma_1$ is concerned model A predicts 
a maximum of 12 mb at the beam
energy $T_{lab}\approx 20$ MeV whereas model D yields a maximum of
practically the same magnitude at $T_{lab}\approx 10$ MeV.
In both cases $\sigma_1$ becomes large and
negative at very low energies due to the dominance of the
Coulomb-nuclear interference term in this region.
For comparison, I include here also results based on the Nijmegen
$\bar pp$ PWA, whose amplitudes were recently re-constructed by
Dmitriev {\it et al.} \cite{Salnikov}. Note that the displayed curves are 
those for $\theta_{acc} = 10$ mrad \cite{Salnikov}.
The results of Ref. \cite{Salnikov} suggest significantly
larger values for $\sigma_1$ over the whole considered energy range,
cf. Fig.~\ref{totppC}.  
Finally, there is also an investigation where a version of the Paris 
$\bar NN$ model \cite{Paris} was employed \cite{DmitrievMS}. In that 
calculation the largest value for $\sigma_1$ was found to be -15 mb 
at $T_{lab}=45$ MeV.

With regard to $\sigma_2$ model A and D predict values around 10 mb
for $\bar pp$ scattering at higher energies. Close to threshold
large negative values are predicted
for $\sigma_2^h+\sigma_2^{int}$ due to the Coulomb-nuclear
interference term. One should note, however, that for beam energies below
5 MeV, say, the total Coulomb cross section becomes very
large. In this case the beam lifetime turns out to be too short and
the spin-filtering method cannot be used for polarization buildup in
a storage ring.
 
The results for $\sigma_2$ based on the Nijmegen $\bar pp$ PWA 
\cite{Salnikov} turn out to be fairly similar to the predictions
of the J\"ulich models, in particular for energies above 100 MeV.
(Please note that our definition for
the cross section $\sigma_2$ differs from that in Ref.~\cite{MS}: our $\sigma_2$
is equal to $\sigma_2-\sigma_1$ as definined in Eq.~(2) of Ref.~\cite{MS}.)
The results based on the Paris $\bar pp$ potential are similar to those
of model D, at least for the energy range covered in Ref.~\cite{DmitrievMS}.

\section{Conclusions}

In this contribution I have reviewed predictions for the 
spin-dependent cross sections $\sigma_1$ and  $\sigma_2$ using either
$\bar NN$ potential models \cite{Hippchen,Mull,Paris} or $\bar NN$
amplitudes from a partial-wave analysis \cite{Timmermans}. There 
are significant differences in the various results -- which is 
certainly not surprising given our incomplete knowledge of the 
spin dependence of the $\bar NN$ interaction. 

The polarization buildup due to the spin-filtering mechanism is determined 
mainly by the ratio of the polarized total cross section $\sigma_i$ (i=1,2) 
to the unpolarized one ($\sigma_0$) \cite{MS}.
For the ratio $\sigma_2/\sigma_0$ all considered interactions predict 
values around 10\% for beam energies above $50$ MeV. 
For $\sigma_1/\sigma_0$ only the Nijmegen PWA yields a ratio of 
comparable magnitude while the potential models predict significantly
smaller values, specifically for higher energies.  
It should be said that yields of around 10\% would be sufficient for the 
requirements of the PAX experiment \cite{Frankpc}.

At present there are plans to investigate the polarization buildup mechanism 
in $\bar p^1$H scattering in a new experiment 
at CERN \cite{AD,AD1}. The stored antiprotons will be scattered off
a polarized $^1$H target in that experiment \cite{AD,AD1} and the polarization
of the antiproton beam will be measured at intermediate energies.
Besides of being a test for the feasibility of spin-filtering for 
antiprotons this experiment will also provide data for the spin-dependent
cross sections $\sigma_1$ and  $\sigma_2$. Such data will be very 
useful for constraining the spin dependence of the $\bar NN$ 
interaction. 


\end{document}